\documentclass[11pt,twocolumn]{article}

\usepackage{fullpage}
\usepackage{times}
\usepackage{algorithm}
\usepackage{algorithmic}
\usepackage{amsthm}
\usepackage{graphicx}

\newtheorem{thm}{Theorem}[section]
\newtheorem{lem}{Lemma}[section]

\normalsize

\newcommand{\RelaxFloats}{
        \renewcommand{\topfraction}{0.9}
        \renewcommand{\floatpagefraction}{0.9}
        \renewcommand{\textfraction}{0.1}
}

\begin{document}

\RelaxFloats

\begin{titlepage}

\flushleft

\vspace{0.00in}
\parbox{6.5in}{\large \noindent
   Draft prepared for \textbf{arXiv}.
}

\vspace{0.10in}
\parbox{6.5in}{\large \noindent
   Manuscript information:
   {10} text pages,
   {2} figures,
   {3} tables.
}

\vspace{2.00in}
\parbox{6.5in}{\LARGE \centering
   Mining Mass Spectra: Metric Embeddings and \\
   Fast Near Neighbor Search
}

\vspace{0.5in}
\parbox{6.5in}{\large \centering
   Debojyoti Dutta,
   Ting Chen\footnotemark[1]
}

\vspace{0.5in}
\parbox{6.5in}{\large \centering
   Molecular and Computational Biology Program \\
   University of Southern California \\
   Los Angeles, CA 90089-2910
}

\vspace{0.5in}
\parbox{6.5in}{\large \centering
   \today
}

\footnotetext[1]{
To whom correspondence should be addressed.
Molecular and Computational Biology Program,
University of Southern California.
MCB 201, 1050 Childs Way,
Los Angeles, CA 90089-2910.
E-mail: ddutta@usc.edu.,
       tingchen@usc.edu
Tel:   (213)740-2416,
      (213)740-2415.
Fax:   (213)740-8631.
}
\end{titlepage}

\normalsize

\begin{small}
\begin{abstract}

Mining large-scale high-throughput tandem mass spectrometry data sets
is a very important problem in mass spectrometry based protein
identification.
One of the fundamental problems in large scale mining of spectra is to
design appropriate metrics and algorithms to avoid all-pair-wise
comparisons of spectra.  In this paper, we present a general framework
based on vector spaces to avoid pair-wise comparisons.
We first robustly embed spectra in a high dimensional space in a novel
fashion and then apply fast approximate near neighbor algorithms for
tasks such as constructing filters for database search, indexing and
similarity searching.  We formally prove that our embedding has low
distortion compared to the cosine similarity, and, along with locality
sensitive hashing (LSH), we design filters for database search that
can filter out more than 989\% of peptides (118 times less)
while missing at most 0.29\%
of the correct sequences. We then show how our framework can be used
in similarity searching, which can then be used to detect tight
clusters or replicates. On an average, for a cluster size of 16
spectra, LSH only misses 1 spectrum and admits only 1 false spectrum.
In addition, our framework in conjunction with dimension reduction
techniques allow us to visualize large datasets in 2D space. Our
framework also has the potential to embed and compare datasets with
post translation modifications (PTM).

\end{abstract}
\end{small}

\section{Introduction}

Proteomics aims to analyze proteins and peptides expressed by the
dynamic biological processes within
cells~\cite{Pandey00,Aebersold00}. Proteins are responsible for many
inter and intra-cellular activities such as metabolism and cell
signaling where proteins are often modified after
translation within cells~\cite{Mann03NatBiot,Yates95}. 
In the post-genomic era, one
of the most important problems is to characterize the {\em proteome},
i.e. the set of proteins within an organism.

Tandem mass spectrometry is one of the most promising and widely used
high throughput techniques to analyze proteins and peptides
\cite{Pandey00,Aebersold00}. It comprises of two stages. A protein mixture 
is enzymatically digested and separated by HPLC (High Performance 
Liquid Chromatography) before inserting into a mass spectrometer through 
a capillary. Then the peptides gets ionized 
and their precursor ion masses, or 
mass/charge ratios,  are measured. This is the MS1
stage. The peaks (or ionized peptides) from the MS1 stage are
selected and further fragmented in a second stage using techniques
such as Collision Induced Dissociation (CID) to yield the MS2 fragment
ions. Ideally, each peptide gets cleaved into two parts. The N-terminal
ion (b-ion) represents the prefix while the C-terminal ion (y-ion) is the suffix.  
This stage is also known as the tandem MS or the MSMS
stage. For more details beyond this oversimplified description, the
reader is directed to the wonderful survey~\cite{Aebersold00}.

There are two main approaches to analyzing tandem mass spectra
data. First, and the most widely used, is the 
database search method~\cite{Keller02AC,Nsvski03,Zhang,Bafna01}.  
Here, peptides from a sequence database are digested in-silico and the
resultant virtual spectra are matched (or scored) with the real
spectra. High scored peptides are typically chosen as the peptide
candidates.  This method leads to a combinatorial explosion
when used to search for Post Translational Modifications
(PTMs)~\cite{Yates95}.  Second, the de-novo
method \cite{Dancik99,Chen01,Ma} reconstructs the sequence without the
help of a database.  
Other approaches combine denovo sequencing and database search by 
first generating sequence tags, or subsequences, and then using these
tags~\cite{pepnovo} as filters for database search with and 
without PTMs~\cite{inspect}.

The promise of tandem mass spectrometry has led research groups to
routinely use this method to probe the proteomes. 
A single run of a mass spectrometer can generate several thousands of
spectra, and the sheer size as well as the number of real life mass
spectra datasets is predicted to grow at an unprecedented rate with
laboratories operating several spectrometers in parallel, round the
clock. Thus, efficient mining of these large-scale mass spectra data
to obtain useful clues for biological discovery is a very important
problem.

Mining large spectra has several challenges, some of which 
are presented below. 
1) Indexing huge databases of mass spectra is not
standardized. Commonly used methods use precursion ion mass but this
method has two main problems: i) there can be errors in precursor ion
masses. ii) there may be many spectra (several thousands of them) that
have masses close to each other.
2) It is difficult to search for similar spectra on a large scale
quickly, or in sublinear time. This is a core function used by several
data mining applications.
3) Clustering large databases of spectra is a daunting task. Most
similarity measures proposed in tandem mass spectrometry use pair wise
metrics for similarity. Such pair wise methods lead to an explosion of
similarity calculations, i.e. $O(n^2)$ for a set of $n$ spectra. Thus,
a key open problem is to use methods that avoid the pair-wise
similarity calculations.  If objects can be transformed into metric
spaces, problems such as similarity searching and clustering becomes
easier.  Thus we need to find methods to robustly embed spectra in
metric spaces.
4) Visualization of large groups of mass spectra is an important
problem which can also be used to qualitatively identify outliers in
the huge number of spectra produced.

In this paper, we present a general framework for large scale mining of 
tandem mass spectra. Our main contributions are the following: 
1) We robustly embed spectra into a metric space,
2) We show, both formally and empirically that distances using our
embedding areas good as those that use the well known cosine method.
3) Then we use apply a geometric fast near neighbor search technique,
Locality Sensitive Hashing (LSH)~\cite{datar04}, to solve several
problems such as fast filters for database search, similarity
searching of mass spectra, and visualization of large spectral
database.
4) Our embedding in conjunction with PCA and manifold learning can be
used to visualize large groups of spectra.
5) Our embedding holds promise for comparing spectra with Post
Translational Modifications (PTM).

Our idea of robust embedding of vector spaces to mine mass spectra is
novel. Previous work to embed spectra into vector spaces using vectors
of amino acid counts to database search~\cite{Halligan04,Halligan05}.
They focussed on clustering sequence databases based on this amino
acid counts to search for mass spectra, given amino acid counts or
sequence tags. However getting an accurate estimate of amino acid
composition is itself a hard problem, especially when the quality of
spectra is not high. However, our method embeds ion fragments of
spectra directly into a vector space and avoids estimating higher
level features such as amino acid composition. Also our scheme is more
general: using a single embedding, we can either compare spectra with
each other or compare spectra with peptide sequences by generating
their virtual, or in-silico digested, spectra.
In addition, we demonstrate that our framework can be used in concrete
mining applications.  We first use our embedding along with Locality
Sensitive Hashing to speed-up database search.  We demonstrate that we
can filter out more than 99.152\% spectra with a false negative rate
of 0.29\%. The average query time for a spectra is 0.21s.
Then, we answer similarity queries and find replicates or tight
clusters.  LSH misses an average of 1 spectrum per cluster, that have
an average cluster size of 16 spectra, while admitting only 1 false
spectrum.  

To the best of our knowledge, we are not aware of any other work that robustly 
embeds spectra in metric spaces with provable guarantees and then uses fast 
approximate near neighbor techniques to solve mass spectrometry data mining 
problems.

\section{Methods}

Our approach is to use vector spaces which have been successful in
numerous data mining applications including web searching{\bf cite web
mining}.  Several fast mining algorithms become simpler to design in
these spaces, compared to designing them in non metric spaces
e.g. spaces where the only available measure is a pairwise similarity
measure.  Thus, the key problem in this approach is to robustly embed
spectra into a high dimensional metric space and define appropriate
distances. Also, these distances must be correlated with the well
known cosine similarities. In other words, we desire an embedding with
bounded distortion with respect to the cosine similarity.

\subsection{Embedding Spectra}

\subsection*{Noise Removal}

\begin{figure}\label{fig:SNRdist}
    \includegraphics[width=\linewidth,height=3in]{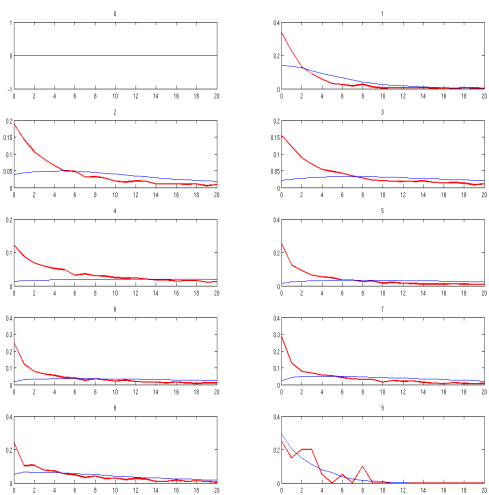}
    \caption{Signal and Noise distributions of peak intensities in different 
regions of spectra (from the training set).}
\end{figure}

The achiles heel of tandem mass spectra analysis is the amount of
noise in the mass spectra. In fact, most peaks (around 80\%) cannot be
explained and are called {\em 'noise'} peaks. {\em 'Signal'} peaks
(such as $b, y$ ions) are useful for interpretation.  As a first step,
we remove noise peaks enriching the signal to noise ratio (SNR).

We use a statistical method to increase SNR.  We first find the
intensity distributions of signal and noise peaks in a set of
annotated spectra.  For this, we consider a set of good quality
annotated spectra as described in Section~\ref{sec:results} and
generate the virtual spectrum $v_p$ for each of the real spectra $r_p$
for a peptide $p$.  For the virtual spectrum generation we consider
the following ions: $b$, $b-H_2O$, $b-NH_3$, $y$, $y-H_20$, $y-NH_3$.
Then we divide the mass range of $r_p$ into $k=10$ sections. For each
section, and for each real peak, we consider its intensity rank
i.e. the most intense peak has rank 0 and so on.  We divide the peaks
of $r_p$ into two sets $S_p$ and $N_p$. $S_p$ contains all those
peaks, and their intensity ranks, which have a match in the virtual
spectrum $v_p$. Thus, for each region, we can get a distribution of
signal and noise intensity ranks for each region as shown in
Figure~\ref{fig:SNRdist}.

We define a  metric SNR of a peak $(mz_j,I_j)$ as  follows 
$$ 
SNR(j)\ =\ \frac{P[\mbox{rank}(j)|(mz_j,I_j)\in S_p]}{P[\mbox{rank}(j)|(mz_j,I_j)\in N_p]}
$$
If larger SNR, the peak is likely to be a useful peak, else its a noise peak.  
From Figure~\ref{fig:SNRdist} we can conclude 
that the noise is very poor at the ends of the spectra, i.e. at low mass
regions and high mass regions. This statistical observation reinforces
the mass spectrometry folklore that the {\em middle region} is the most 
suitable for finding signal peaks.

\subsection*{Features and Distances}

There are several possible ways to embed tandem mass spectra into a
vector space that support the most common operation of comparing two
spectra and find similarities. For example, the cosine similarity
metric~\cite{Keller02AC} and their different variants have been very
popular in the recent papers.  Unfortunately the cosine metric does
not yield a metric embedding because the triangle inequality is
violated.  Also the cosine similarity metric implies algorithms that
consider pairs of spectra. Clearly such algorithms are difficult to
scale due to the $O(n^2)$ number of similarity calculations.

For metric embeddings, the design space is quite large. 
A simple idea is to directly bin the peaks and use the intensities to
form a vector space. However spectra from different datasets have
different intensities and we would like to have a single embedding
that could potentially integrate multiple spectral databases.

\begin{figure}\label{fig:cube}
\begin{tabular}{c c}
\includegraphics[width=1.5in,height=1.5in]{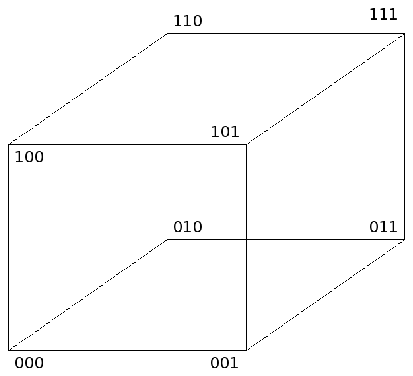} &
\includegraphics[width=1.5in,height=1.5in]{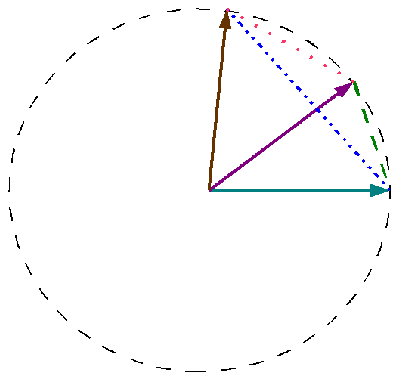}\\
(i) & (ii)\\
\end{tabular}
\caption{ (i) Embedding spectra in a $n$-dimensional cube, (ii) 
Using a 2-dimensional example to illustrate the correlation between the 
Euclidean distance and the well known cosine similarity }
\end{figure}

We first {\em clean} spectra as mentioned in the previous subsection.
Then we divide the entire mass range (from 0 to some maximum range)
into discrete intervals of 2da.  For each interval of 2da,
a bit is set to 1 if the cleaned spectrum
contains a peak in that interval, else it is 0. This embeds
each spectra into the vertices of a n-dimensional hypercube. A 3D
version is shown in Figure~\ref{fig:cube}. Our feature vectors
are defined to be the the {\em unit} vectors in the direction of the
corresponding vertices of the n-dimensional hypercube. Thus the space
of our embedding is a n-dimensional unit hyper-sphere.

We define the spectral similarity or distance between spectra 
$x$, $y$, as $||x-y||$. If the angle between two similar spectra $x$, $y$ is
$\theta$, $\cos \theta$ will be close to 1, or $1-\cos \theta$ will be
very small. Since $x$ $y$ are unit vectors, their
Euclidean distance will also be small.  Thus, for small angles, $1-\cos
\theta \approx D(x,y) $, where $D$ is the Euclidean distance. It is easy
to show that as $n$ or the number of dimensions increases, the minimum
angle for pairs of very similar spectra $x$, $y$ becomes
smaller. Thus, instead of calculating the $1- \cos \theta$, we 
calculate $D(x,y)$.  The natural question that arises is the
{\em distortion} of our embedding.  We will now show that it is has
bounded accuracy in theory, and we will later show that the accuracy
is empirically quite high in comparision with the cosine similarity.

We prove some properties of the embeddings. It is easy to show the following theorem:
\begin{thm}
The embedding discussed above defines a metric space. 
\end{thm}
\begin{proof}
The proof is very simple 
To show that our embedding defines a metric space, we need to prove three things:
1) $||x-y||=0$ iff $x=y$, 2) $||x-y||=||y-x||$ and 3) the distance measure 
obeys the triangle inequality. These properties are trivial to prove in our case 
as our embedding uses Eculidean distances. 
\end{proof}

We then  show that the maximum euclidean distance is bounded by $\sqrt{2}$. 
\begin{lem}
The distance between the feature vectors of any two mass spectra is
bounded above by $\sqrt{2}$.
\end{lem}
\begin{proof}
Suppose there are two spectra $x$, $y$ respectively. We shall uses the
names of the spectra and their feature vectors
interchangeably. According to our scheme we first filter the noisy
peaks and generate the binary vector after binning. Now assume $x$ has
$k$ bits set to a and $y$ has $k'$ bits set to 1. Also assume that $c$
of the common bits are 1.  Then $||x||=\frac{1}{\sqrt{k}}$ and
$||y||=\frac{1}{\sqrt{k'}}$. Since $c$ bits are common, the number of
dissimilar bits between $x$ and $y$ are $(k-c)+(k'-c)$. We have
\begin{small}
\begin{eqnarray}
||x-y|| &  = &  \sqrt{(k-c).\left( \frac{1}{\sqrt{k}}\right)^2\ +
  \ (k'-c).\left( \frac{1}{\sqrt{k'}}\right)^2 }\\
 &  = & \sqrt{ 2 -  c.\left( \frac{1}{k} + \frac{1}{k'} \right) }
\end{eqnarray}
\end{small}
\end{proof}

Next we show that our embedding has bounded distortion when we compare
with the well known cosine similarity.  We have the following theorem:
\begin{thm}
If $\theta$ is the angle made by the feature vectors of spectra $x$,
$y$, and the number of ones in each of the vectors after binning is
the same we must have
$0<\frac{1-\cos{\theta}}{||x-y||}<\frac{1}{\sqrt{2}}$. Or in other
words, the distortion between our Euclidean embedding and the cosine
similarity is bounded.
\end{thm}
\begin{proof}
As in the previous lemma, 
$
||x-y|| = \sqrt{ 2 -  c.\left( \frac{1}{k} + \frac{1}{k'} \right) }.
$
Now the cosine of the angle $\theta$ between $x$, $y$ can be written as 
$\cos\theta = \frac{c}{\sqrt{kk'}}$.
Assume $k=k'$ and note that $0\leq \frac{c}{k}\leq 1$. Thus, we must have 
\begin{eqnarray}
\frac { 1 - \cos\theta } {||x-y||} &  =  & 
\frac { 1\ -\ \frac{c}{\sqrt{kk'}} }
{ \sqrt{ 2 -  c.\left( \frac{1}{k} + \frac{1}{k'} \right) } } \\
 & = & \frac{1-\frac{c}{k}}{\sqrt{2-\frac{2c}{k}}}\\
 & = & \frac{1}{\sqrt{2}} \sqrt{1-\frac{c}{k}}
\end{eqnarray}
We note that since,  $0\leq \frac{c}{k}\leq 1$, we must also have 
 $0\leq 1- \frac{c}{k}\leq 1$ and the theorem follows. 
\end{proof}

Thus our embedding will perform almost as good as the standard cosine
metric. We show in the next section that this is indeed the case,
empirically.  Also, since the points are in a Euclidean space, we can
elegant geometric techniques that yield fast approximate algorithms
for mining the data.

\subsection{Similarity Searching}

The ability to calculate distances as opposed to cosines is an
important feature of our framework. Now, we apply elegant 
near neighbor algorithms to answer queries quickly but
approximately, as we show in the paper.  The basic query
primitive we use is the following:

\noindent{Primitive 1}: Given a spectrum $x$ and a set of spectra $S$,
we want to find all the spectra $S_r$ that are similar to $x$,
i.e. spectrum $y\in S_r$, iff $D(x,y)<r_q$, where $D$ is the Euclidean
distance and the $r_q$ is a query radius.

A very simple approach would be to do a linear scan on the database
and output every spectrum $y$ such that $D(x,y)<r_q$. This takes
$O(n)$ time. However, if $S$ becomes very large and so do the number
of queries say $O(n)$, then we have a $O(n^2)$ algorithm. This is
clearly unacceptable for our problem. Thus, we desire methods that
will yield near neighbor queries in {\em sub-linear} time. For this we
are willing to tradeoff some accuracy for speedup. Several sub-linear
near neighbor methods exist but we leverage Locality Sensitive
Hashing~\cite{datar04} since, unlike others, it promises bounded
guarantees and is also easy to implement.  We briefly present the idea
below.

\subsection*{Locality Sensitive Hashing}

The basic idea behind random projections is a class of hash functions
that are locality sensitive i.e. if two points $(p, q)$ are close they
will have small $|p-q|$ and they will hash to the same value with high
probability. If they are far they should collide with small
probability.

\noindent{Definition 1}: A family $\{ H = f: S \rightarrow U \}$ is
called locality-sensitive, if for any point $q$, the function $$p(t) =
Pr_H[h(q) = h(v) : |q-v| = t]$$ is strictly decreasing in $t$. That
is, the probability of collision of points $q$ and $v$ is decreasing
with the distance between them.

\noindent{Definition 2}: A family $H=\{h:S\rightarrow U\}$ is called
$(r_1,r_2,p_1,p_2)$ sensitive for distribution $D$ if for any $v,q \in S$,
we have
\begin{itemize}
\item if $v\in B(q,r_1)$ then $\mbox{Pr}[h(q)=h(v)]\geq p_1$
\item if $v\notin B(q,r_2)$ then $\mbox{Pr}[h(q)=h(v)]\leq p_2$
\end{itemize}
Here $B(q,r)$ represents a ball around point $q$ with a radius $r$.
Thus a good family of hash functions will try to {\em amplify}
the gap between $p_1$ and $p_2$.

Indyk et.~al.~\cite{datar04} showed that s-stable distributions can be
used to construct such families of locality sensitive hash
functions. An s-stable distribution is defined as follows.

\noindent{Definition 3}: A distribution $D$ over $R$ is called {\em
s-stable}, if there exists $s$ such that for any $n$ real numbers $v_1
... v_n$ and i.i.d. variables $X_1 ... X_n$ with distribution $D$, the
random variable $\sum_i{v_i X_i}$ has the same distribution as the
variable $(\sum_i{v_i^p})^{\frac{1}{s}}X$, where $X$ is a random
variable with distribution $D$.

Consider a random vector $a$ of $n$ dimensions. For any two
n-dimensional vectors $(p, q)$ the distance between their projections
$(a.p - a.q)$ is distributed as $|p-q|_s X$ where $X$ is a s-stable
distribution. We {\em chop} the real line into equal width segments of
appropriate size and assign hash values to vectors based on which
segment they project onto. The above can be shown to be locality
preserving.

There are two parameters to tune LSH. Given a family $H$ of hash
functions as defined above, the LSH algorithm chooses $k$ of them and
concatenates them to amplify the gap between $p_1$ and $p_2$. Thus,
for a point $v$, $g(v)=(h_1(v)...h_k(v))$. Also, $L$ such groups of
hash functions are chosen, independently and uniformly at random,
(i.e. $g_1...g_L$) to reduce the error.  During pre-processing, each
point $v$ is hashed by the $L$ functions buckets and stored in the
bucket given by each of $g_i(v)$. For any query point $q$, all the
buckets $g_1(q)...g_L(q)$ are searched. For each point $x$ in the
buckets, if the distance between $q$ and $x$ is within the query
distace, we output this as the nearest neighbor. Thus, the parameters
$k$ and $L$ are crucial.  It has been shown~\cite{indyk99,datar04}
that $k=\log_{1/p_2}{n}$ and $L=n^\rho$, where
$\rho=\frac{\log{1/p_1}}{\log{1/p_2}}$, ensures locality sensitive
properties. In Ref.~\cite{datar04}, the authors consider $L2$ spaces
and bound $\rho$ above empirically by $\frac{1}{c}$, $c$ being the
approximation guarantee, i.e. for a given radius $R$, the algorithm
returns points whose distance is within $c\times R$.  The time
complexity of LSH has been shown to be $O(dn^\rho \log{n})$, where $d$
is the number of dimensions and $\rho$ is as defined above.  Thus, if
we desire a coarse level of approximation, LSH can guarantee
sub-linear run times for geometric queries.

\subsection{Similarity Searching}

Using our embedding and a fast near neighbor algorithm, we can 
find spectra similar to a given query spectrum. The
key is to use the correct query radius $r$. We
show in the next section how this can be chosen. 
If we give too high a radius, it might yield a
large dataset and if the radius is too low, it might not yield any
neighbor.

If an appropriate query radius is chosen, 
it is easy to find tight clusters using the following heuristic:
\noindent{ANN-cluster}: 1) Embed spectra into a Euclidean space and
form the set $S$.  2) Hash the feature vectors, $S$, using LSH. 3)
Choose some $k$ random spectra, find their near neighbors (tight
clusters). For each random spectra add their neighbors to set $S$. 4)
$S=S-C$. 5) Go to step 3 till $S$ is empty.

Another immediate consequence of our framework is to find outliers. To
check for outlier, we need to determine whether a spectrum has at most
1 or 2 neighbors. If the neighbors remain unchanged even on increasing
the query radius by $\delta$, a spectrum is indeed an outlier.  Since
near neighbors take sub-linear time with LSH, outliers can be detected
in sub-quadratic time.

\subsection{Speedup Database search}

In this section, we discuss a sample application using our mining
framework.  Database search is the primary tandem mass spectrometry
data mining applications. Given a query spectrum $x$, and a mass
spectra database $MSDB$ (described in Section~\ref{sec:results}, the
problem is to find out which peptide $p\in MSDB$ corresponds to $x$.

Database search is a well explored topic, see ~\cite{Wan05} for
example. Most tools index the the MSDB by the peptide mass. Then for a
spectrum $x$, the precursor mass $m_x$ is found. Then all the spectra
$S_p={y|y\in MSDB}$ are compared with $x$ such that $|m_y-m_x|<\delta$,
where $\delta$ is some pre-defined mass tolerance.  Each comparison
operation between the query spectrum and the candidate spectrum takes
a while depending on the scoring function used.  We reduce the size of
$S_p$ by filtering the unrelated spectra, speeding up the search. We
ensure that we do not filter out the true peptide for a
given spectrum while we discard most of the unrelated peptide.

We generate the virtual spectra from each peptide
sequence in the database, and then embed those virtual spectra in the
Euclidean space, as mentioned. Then for filtering, we choose an appropriate
threshold radius $r$ and query the LSH algorithm to yield all the
candidates within a ball of radius $r$. The ratio of the total number
of peptides within a mass tolerance divided by the number of
candidates returned is our speedup.

\subsection{Visualization and Dimension Reduction}

As mentioned earlier, vizualizing thousands of spectra is a very hard
problem.  We are not aware of any previous work that allows us to
visualize large mass spectrometry data sets.  Our embedding followed
by dimension reduction allows to view spectra on a two or three
dimensional space. As a bonus, it qualitatively allows us to identify
outliers in the data set.

Once we have embedded the spectra in a Euclidean space, we can use
some of the common techniques to visualize high dimensional data by
dimensionality reduction.  The most common linear method is to use
PCA~\cite{strang}. Recently, several non-linear methods for
dimensionality reduction have been discovered, the majority of them
exploiting the low dimensional manifold structure of the dataset.  In
this paper, we leverage one of these techniques, the isomap method, to
project the high dimensional data on a 2D plane. Due to lack of space
we do not provide a description of the method.

\section{Experimental Results}
\label{sec:results}

In this section, we describe the empirical evaluation of our embedding
followed by some representative data mining tasks. Unless otherwise
stated we use the following dataset from Keller
et. al.~\cite{Keller02Data}. For calculating statistics, we used 80\%
of the 1618 spectra from this annotation at random.  The statistics
were independent of the exact choices of the spectra.  Note that our
techniques are unsupervised except for the selection of query radii.
Out of this, 1014 spectra were digested with trypsin and were used for
database search filter.

For database search filters, a non-redundant protein sequence database
called MSDB, which is maintained by the Imperial College, London.  The
release (20042301) has 1,454,651 protein sequences (around 550M amino
acids) from multiple organisms.  Peptide sequences were generated by
in-silico digestion and the list of peptides were grouped into
different files by their precursor ion mass, a different file for
10da.

\subsection{Empirical evaluation of the embedding}

\begin{figure}\label{fig:inter}
    \includegraphics[width=\linewidth,height=4in]{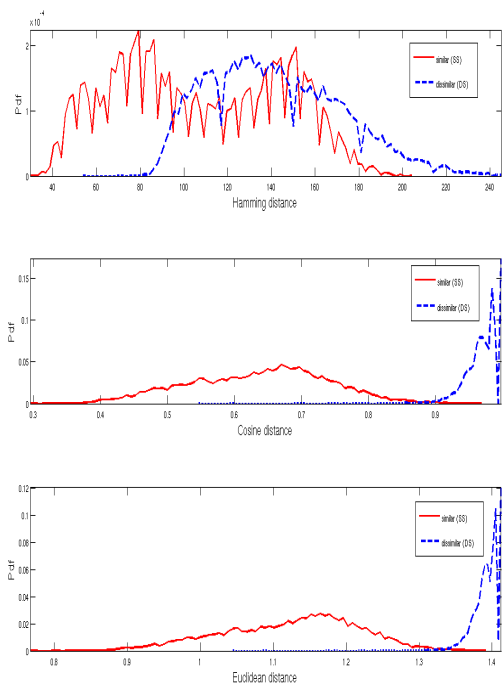}
    \caption{Distribution of scores with real spectra using different
metrics (hamming, 1-cosine, euclidean). The dotted curve plots the
inter-cluster distances while the solid line represents the
intra-cluster distribution.}
\end{figure}

In this section, we critically analyze our embedding and different
distance metric. For these analyzes, we chose a set of 1014 curated
spectra of proteins digested with trypsin and reported by Keller
et. al.  We then cleaned the spectra picked the most likely to be the
signal peaks. Then we constructed the binary bit vector as discussed
earlier. For the set of spectra, we knew that there were 100 odd
clusters with 15 spectra per cluster on an average.  We calculate the
pairwise distances between spectra within the same cluster and we term
this the similar set $SS$.  We then choose a representative from each
cluster at random and calculate the distances and we call this set the
dissimilar set $DS$.  Then we plot the frequency distribution of DS
and SS as they both have similar number of pairwise distances in
Figure~\ref{fig:inter} for three metrics: hamming, 1-cosine and
euclidean. Its very clear that hamming is unsuitable as a metric as it
has low discriminability. As expected, 1-cosine and euclidean looks
almost similar with low overlaps between the sets DS and SS. Also note
that the cosine metric used here is not exactly the same used by
others.  We do not take the intensities into consideration after we
have selected the peaks.

\begin{figure}\label{fig:trueVSfalseDB}
    \includegraphics[width=\linewidth,height=4in]{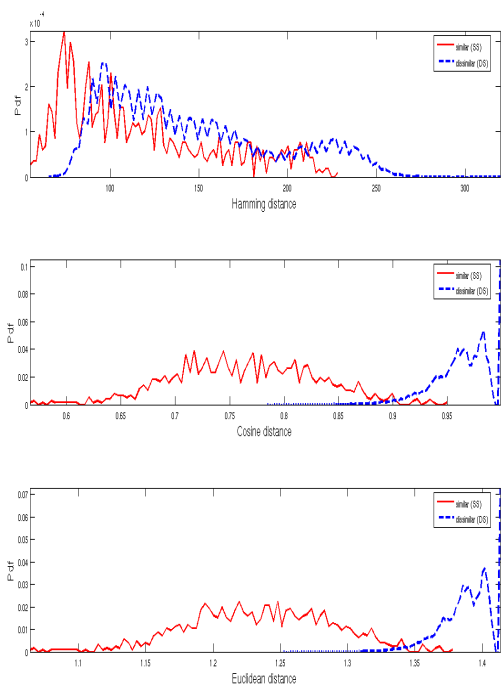}
    \caption{Distribution of distance between real and virtual spectra using 
different metric. The dotten curve represents the distance between 
real spectra and distances to virtual spectra from 100 different peptides
of similar precursor masses. The sequences are from MSDB.  
The other curve shows the distribution of distances between spectra 
and the virtual spectra from the true peptides.}
\end{figure}

Now, we consider the database of tryptic peptides, $MSDB$. For each
peptide, we generate its virtual spectrum and then construct the
feature vector as above.  For each real spectrum, we calculate the
distance with the correct virtual spectra and we call this set of
scores to be $SS$. Then we choose, from the database, 100 random
peptides having almost the same mass as the precursion ion mass of the
given spectrum. We then add the set of scores to the dissimilar set
$DS$. We then plot the probability distribution of SS and DS in
Figure~\ref{fig:trueVSfalseDB}. Again we can see the clear sepatation
between the two sets of distances (with $<1\%$ overlap).  This
indicates that the efficacy of euclidean distance in our embedded
space is a good metric to design filters for database search, Note the
sharp impulse at 1.414 corresponding to distances between real spectra
and completely dissimilar peptides within a mass tolerance of 2da,
providing empirical evidence for Lemma 2.2.

\subsection{Post Translational Modifications}

Now we present some very preliminary results on a set of spectra from
the PFTau protein. We picked 8 good quality spectra with known
Phosphorylations. We wanted to study whether our metric can help
design filters that might work for PTM studies. From the Figure~\ref{fig:ptm},
we note that distances between spectra and their PTM variants 
have a higher likelihood of being classified as similar than dissimilar. 
This is evident from Figure~\ref{fig:inter}.

\begin{figure}\label{fig:ptm}
\begin{small}
\begin{tabular}{ |c | c |} 
\hline
R.LTQAPVPMPDLKNVK.S & 1.23\\ 
R.LTQAPVPMPDLK\# NVK.S & \\
R.HLSNVSSTGSIDMVDSPQLATLADEV & 1.27\\
R.HLSNVSST\^GS\^IDMVDS\^PQLATLADEV & \\
R.TPSLPTPPTR.E & 0.98\\
R.TPSLPT\*PPTR.E &  \\
R.QEFEVMVMEDHAGTYGLGLGDR.K & 1.19\\
R.QEFEVMVMEDHAGT\^YGLGLGDR.K & \\
\hline
\label{tab:ptm}
\end{tabular}
\end{small}
\caption{Some sample distances between spectra and their PTM variants.
Note the low scores between the pairs. Distances between spectra of 
different peptides had a mean $\mu=1.388$ and $\sigma=0.017$.}
\end{figure}

\subsection{Query processing using LSH}

In this section, we quantify the accuracy of our framework for
similarity searching and clustering.  As mentioned earlier, we use LSH
to answer queries with bounded errors in expected sub-linear time.

\begin{figure}\label{fig:LSH-misses}
    \includegraphics[width=\linewidth,height=1.75in]{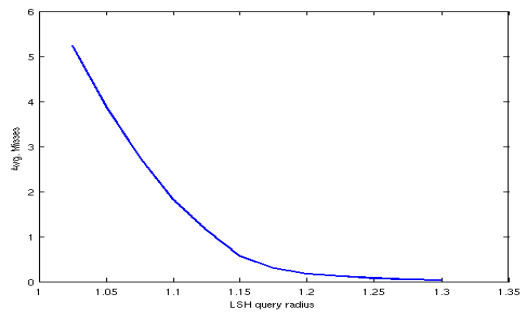}
    \caption{The average number of spectra that are present in the cluster 
containing the query spectrum but are missed by LSH }
\end{figure}

\begin{figure}\label{fig:LSH-fpos}
    \includegraphics[width=\linewidth,height=1.75in]{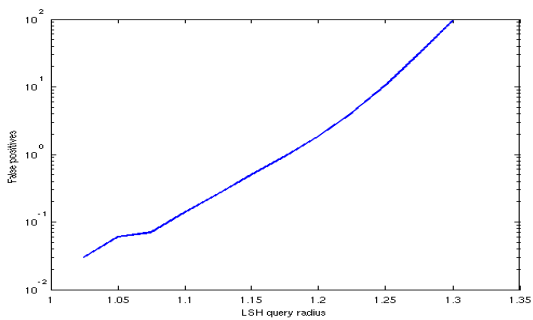}
    \caption{The average number of spectra that are not present in the cluster
containing the query spectrum but are reported by LSH}
\end{figure}

We first indexed the 1014 spectra using our embedding followed by LSH.  
For each of the 1014 spectra, we queried LSH with a radius $r$. 
We varied $r$.  We plot the
number of missed spectra that were actually present in the cluster of
the query spectrum in Figure~\ref{fig:LSH-misses} and the number of
false positives in Figure~\ref{fig:LSH-fpos}.  As we increased the
radius, we the number of misses decreased. This is expected as the
radius of the {\em query ball} increases the number of possible data
points that can be considered. As expected, the number of false
positives also increased as $r$ increased. This indirectly demonstrates
the accuracy of any clustering algorithm based on LSH. We 
miss an average of 1 spectrum within each cluster
while admitting only 1 false spectrum.

At $r=1.0-1.1$ the false positives are not very high.  This might be
important when we want to query for similar spectra in order to
generate the consensus spectra. In such situations, it might be fine
to miss out some bad quality spectra (distances to bad quality spectra
are usually higher). Also, consider situations where we would like to
coarsely partition the data set (e.g. for clustering).  Then, 
we can afford to have a few false positives but we cannot
miss any true positives. In such cases we increase the radius to at most 
1.25 as the likelihood of a intra-cluster distance being greater than 
1.25 is low, from Figure~\ref{fig:inter}.

\subsection{Speeding up Database Search}

\begin{figure}\label{fig:speedup}
    \includegraphics[width=\linewidth,height=2in]{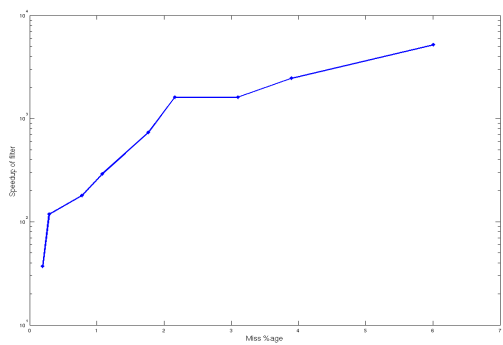}
    \caption{Filtering of spectra for DBASE search}
\end{figure}

To test the efficacy of our framework on speeding up database search,
we first use our metric to filter out candidate spectra. Since our
distance calculation is much faster than the detailed scoring of two
spectra, we define speedup by the ratio of total number of candidate
peptides with a mass tolerance of 2 daltons and the total number of
peptides that have a distance of $\Delta$ with the query spectrum and
have the same mass tolerance. Then we increase $\Delta$ and calculate
the number of true peptides missed in this filtering process.  In
Figure~\ref{fig:speedup} we plot the speedup on a logarithmic scale
against the miss percentage. This gives us the speedup (or quality of
filtering) versus accuracy tradeoff of using our framework.  For a 2
dalton range the number of peptides are around 100-200K. For around a
a 100K peptide set, LSH takes 0.21s on an average to answer queries. 
As we see from Figure~\ref{fig:speedup}, we can get an
average speedup of 118 if we allow 0.19\% misses. 
This may be reasonable for
many applications. In fact, we found that our errors were due to low
quality spectra in our test dataset.

\subsection{Visualization and Dimension Reduction}

\begin{figure}\label{fig:pca}
\begin{tabular}{c c}
\includegraphics[width=1.5in,height=1.75in]{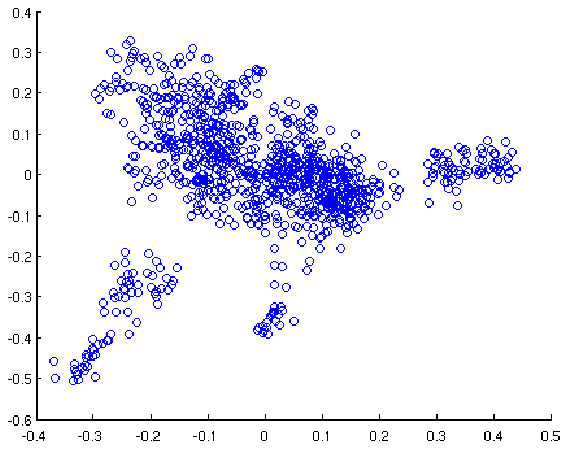} &
\includegraphics[width=1.5in,height=1.75in]{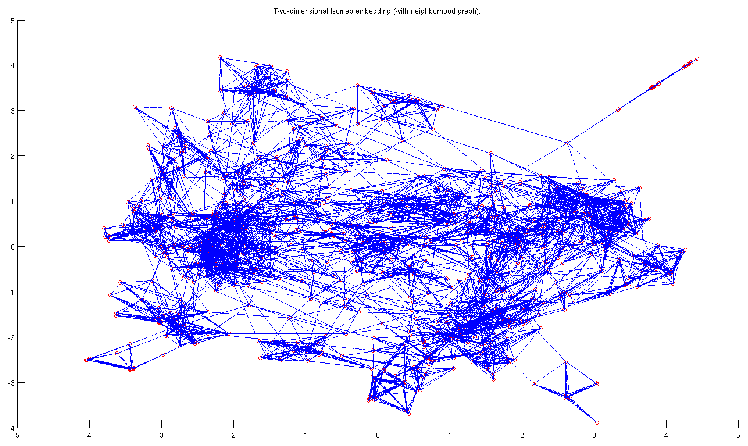}\\
(i) PCA & (ii) Isomap\\
\end{tabular}
    \caption{Dimension Reduction with Isomap}
\end{figure}

Consider the training dataset of mass spectra.  We first generate
Euclidean feature vectors for each spectra.  Then we used PCA and
plotted the first two components on the x-axis and the y-axis as shown
in Figure~\ref{fig:pca}(i). The clusters are visible and so are the
outliers. But the visualization is coarse grained.

Then we use Isomaps on the same dataset.  Recall that in Isomaps, one
first needs to calculate the near neighbors. Thus in our plot, we also
show the near neighbor graph along with the projected points as shown
in Figure~\ref{fig:pca}(ii).  The cluster structure seem to be
qualitatively clearer than with PCA.

\section{Discussion}

The results in the previous section look promising. The clear
separation between the DS and SS set during the metrics comparision
was a surprise to us, initially. One of the reasons for the good
result is the quality of the dataset. We first wanted to validate our
simple assumptions and claims on a dataset which had reliable
interpretations. Since we first transform the spectra into binary bit
strings we avoided the huge variations of density in spectra.  The
signal to noise ratio pilot study also underscored the fact that we
need to study spectra by segmenting them. Note that one reason why we
obtained clear separations between the DS and SS in all cases with our
embedding is that we avoided using precursor ion mass as a feature.
Even though its fine to use the precursor mass as a coarser grain
filter, it will lead to less robust embeddings as such masses are
prone to errors due to isotope effects. Also our theoretical results
will not hold.

For LSH, the speed and the accuracy is quite satisfying.  However,
there are two implementation issues.  Our current indexing is memory
bound. This means we need lots of memory to index millions of mass
spectra. Even though this is possible with the current 64 bit
machines, we need to design disk based LSH schemes.  We are working on
a large scale implementation of our framework based on such
techniques. Another issue is the choice of the number of bins and the
mass coverage. Increasing the number of bins leads us to the curse of
dimensionality which would slow down LSH and reduce the filtering
speedup. If we choose fine grained bins with a lower maximum mass, our
embedding will result in a pseudo-metric space as several different
spectra will now satisfy assumption one in Theorem 2.1.

\section{Conclusions and Future Work}

In this paper, we showed that our embedding with geometric algorithms
provides a good framework for mining mass spectra. In particular, we
have demonstrated both theoretically as well as empirically, that our
embedding coupled with Euclidean distance performs as well as the well
known cosine similarity while providing us with the benefits of a
metric space and enabling us to use approximate sub-linear time near
neighbor techniques for data mining. Using this framework, we showed
how we can do similarity searches and find tight clusters. Also, we
demonstrated that we can get 2 order of magnitude filtering for
database search. As an aside, we are also able to visualize large
datasets in two dimensions qualitatively identifying the outliers.

This work is the first step in the direction of an integrated
framework for large scale mining of tandem mass spectra using simple
techniques from embeddings, vector spaces and computational
geometry. Several directions are being investigated at this point. The
main areas of investigation are 1) Better embeddings that offer better
resolution for PTM spectra 2) Faster external database searching
algorithms that use embedding 3) More effective blind PTM searching
using embeddings 4) Large scale clustering and visualization of mass
spectrometry data and 5) Integrating data from different sources using
our embeddings.

We should note that several sections in the paper could be of
independent interest. For example, we need to explore the
probabilistic cleaning of mass spectra in more details.  Our embedding
promises to work across datasets and this general method can be used
to do integrated study of other biological datasets eg. microarray
data sets.

\bibliographystyle{plain}
\bibliography{msms,lsh}

\begin{thebibliography}{10}

\bibitem{Aebersold00}
R.~Aebersold and M.~Mann.
\newblock Mass spectrometry-based proteomics.
\newblock {\em Nature}, 422,6928 (2003),198-207.

\bibitem{Bafna01}
V.~Bafna and N.~Edwards.
\newblock Scope: a probabilistic model for scoring tandem mass spectra against
  a peptide database.
\newblock {\em Bioinformatics}, 17, Suppl 1 (2001),S13-21.

\bibitem{Chen01}
T.~Chen, M.Y. Kao, M.~Tepel, J.~Rush, and G.~Church.
\newblock A dynamic programming approach to de novo peptide sequencing via
  tandem mass spectrometry.
\newblock {\em J Comput Biol}, 8,(2001),325-37.

\bibitem{Dancik99}
V.~Dancik, T.A. Addona, K.R. Clauser, J.E. Vath, and P.A. Pevzner.
\newblock De novo peptide sequencing via tandem mass spectrometry.
\newblock {\em J Comput Biol}, 6,3-4 (1999),327-42.

\bibitem{datar04}
M.~Datar, N.~Immorlica, P.~Indyk, and V.~S. Mirrokni.
\newblock Locality-sensitive hashing scheme based on p-stable distributions.
\newblock In {\em SCG '04: Proceedings of the twentieth annual symposium on
  Computational geometry}, pages 253--262, New York, NY, USA, 2004. ACM Press.

\bibitem{pepnovo}
A.~Frank, S.~Tanner, V.~Bafna, and P.~Pevzner.
\newblock Peptide sequence tags for fast database search in mass-spectrometry.
\newblock {\em J. Proteome Res.}, 2005; 4(4); 1287-1295.

\bibitem{indyk99}
Aristides Gionis, Piotr Indyk, and Rajeev Motwani.
\newblock Similarity search in high dimensions via hashing.
\newblock In {\em VLDB '99: Proceedings of the 25th International Conference on
  Very Large Data Bases}, pages 518--529, San Francisco, CA, USA, 1999. Morgan
  Kaufmann Publishers Inc.

\bibitem{Halligan04}
B.D. Halligan, E.A. Dratz, X.~Feng, S.N. Twigger, P.J. Tonellato, and A.S..
  Greene.
\newblock Peptide identification using peptide amino acid attributed vectors.
\newblock {\em J. Proteome Res.}, 2004,3,813--820.

\bibitem{Halligan05}
B.D. Halligan, V.~Ruotti, S.N. Twigger, and A.S.. Greene.
\newblock Peptide identification using peptide amino acid attributed vectors.
\newblock {\em Nucleic Acid Research}, 2005,Vol33,WebServer issue.

\bibitem{Keller02AC}
A.~Keller, A.I. Nesvizhskii, E.~Kolker, and R.~Aebersold.
\newblock Empirical statistical model to estimate the accuracy of peptide
  identifications made by ms/ms and database search.
\newblock {\em Anal. Chem.}, 74,20(2002),5383-92.

\bibitem{Keller02Data}
et~al. Keller, A.
\newblock Experimental protein mixture for validating tandem mass spectral
  analysis.
\newblock {\em {OMICS}}, 6,2 (2002),207-12.

\bibitem{Ma}
B.~Ma, A.~Doherty-Kirby, and G.~Lajoie.
\newblock Peaks: powerful software for peptide de novo sequencing by tandem
  mass spectrometry.
\newblock {\em Rapid Commun. Mass. Spectrometry}, 17,20 (2003),2337-42.

\bibitem{Mann03NatBiot}
M.~Mann and O.~Jensen.
\newblock Proteomic analysis of post-transaltional modifications.
\newblock {\em Nature Biotechnology}, 21,255-261, 2003.

\bibitem{Nsvski03}
A.~Nesvizhskii, A.~Keller, E.~Kolker, and R.~Aebersold.
\newblock A statistical model for identifying proteins by tandem mass
  spectrometry.
\newblock {\em Analytical Chemistry}, 75, 4646-4658, 2003.

\bibitem{Pandey00}
A.~Pandey and M.~Mann.
\newblock Proteomics to study genes and genomes.
\newblock {\em Nature}, 405, 837-846, 2003.

\bibitem{strang}
G.~Strang.
\newblock {\em Linear Algebra}.
\newblock 3rd edition, 2003.

\bibitem{inspect}
S.~Tanner, H.~Shu, A.~Frank, L.C Wang, E.~Zandi, M.~Mumbi, P.~Pevzner, and
  V.~Bafna.
\newblock {INSPECT}: Identification of posttranslationally modified peptides
  from tandem mass spectra.
\newblock {\em Anal.Chem.}, 77(14) pp 4626 - 4639,2005.

\bibitem{Wan05}
Y.~Wan and T.~Chen.
\newblock A hidden markov model based scoring function for tandem mass
  spectrometry.
\newblock In {\em RECOMB 2005}.

\bibitem{Yates95}
J.R. Yates~3rd., J.K. Eng, A.L. McCormack, and D.~Schieltz.
\newblock Method to correlate tandem mass spectra of modified peptides to amino
  acid sequences in the protein database.
\newblock {\em Anal. Chem.}, Apr 15;67(8):1426-36, 1995.

\bibitem{Zhang}
N.~Zhang~et.al.
\newblock Probid: a probabilistic algorithm to identify peptides through
  sequence database searching using tandem mass spectral data.
\newblock {\em Proteomics}, 2,10 (2002),1406-12.

\end{thebibliography}

\end{document}